# Rethinking Software Engineering in the Foundation Model Era: From Task-Driven AI Copilots to Goal-Driven AI Pair Programmers

Ahmed E. Hassan°, Gustavo A. Oliva★, Dayi Lin★, Boyuan Chen★, Zhen Ming (Jack) Jiang•

cse@huawei.com

★ Centre for Software Excellence, Huawei Canada ° Queen's University, Canada • York University, Canada

## ABSTRACT

The advent of Foundation Models (FMs) and AI-powered copilots has transformed the landscape of software development, offering unprecedented code completion capabilities and enhancing developer productivity. However, the current task-driven nature of these copilots falls short in addressing the broader goals and complexities inherent in software engineering (SE). In this paper, we propose a paradigm shift towards goal-driven AI-powered pair programmers that collaborate with human developers in a more holistic and context-aware manner. We envision AI pair programmers that are goal-driven, human partners, SE-aware, and self-learning. These AI partners engage in iterative, conversation-driven development processes, aligning closely with human goals and facilitating informed decision-making. We discuss the desired attributes of such AI pair programmers and outline key challenges that must be addressed to realize this vision. Ultimately, our work represents a shift from AI-augmented SE to AI-transformed SE by replacing code completion with a collaborative partnership between humans and AI that enhances both productivity and software quality.

## 1 INTRODUCTION

Copilots took developers by storm. Copilots are advanced code completion systems that are powered by Foundation Models (FMs). Copilot solutions such as GitHub Copilot [37] are currently integrated into all popular IDEs and are used by millions of developers all over the world. Research suggests that GitHub Copilot may have a beneficial effect on developer productivity, particularly on mundane repetitive tasks [17, 42, 54].

Despite the power and value of copilots, they are far from being a silver bullet. The hype around novel technologies such as copilots have systematically created inflated and often unrealistic expectations [21]. A closer look at copilots reveal the following limitations:

• **Additive in nature.** Copilots such as GitHub Copilot can only *add* code. This is a major flaw from a SE standpoint, since *adding code* is not always the best solution to a problem. In fact, senior software engineers will frequently leverage design patterns and refactoring techniques to *shorten* source code by creating abstractions that encapsulate concerns and promote reuse. In practice, more code implies more spots for bugs to be found and a harder-to-understand system.

• **Little automation.** Complex software systems are made up of several pieces that are interconnected according to some architectural style and design philosophy. Each system also typically has its own coding conventions, license use policies, and set of dependencies. Yet, copilots have little to no knowledge about all of these essential characteristics while suggesting changes. For instance, GitHub Copilot only knows about the currently opened file as well as opened tabs in the IDE. It is thus up to the developer to coordinate (potentially many) copilot assists and ensure that the resulting code is consistent with the code base. This lack of proper automation contributes to low developer productivity and raises trustworthiness concerns.

• **Problems with the generated code.** Recent research has shown that code produced by GitHub Copilot may be buggy [38] and have performance issues [18]. Indeed, research from Microsoft itself points that senior programmers are less likely to accept GitHub suggestions compared to junior programmers [17, 42, 54].

• **Difficulty in communicating with FMs.** Developers complain about inability to control copilots and make sense of long lists of suggestions (e.g., code completion alternatives) [3].

The aforementioned limitations call for a deep rethinking of the ways in which we have leveraged AI to engineer software systems. In our vision, the fundamental flaw of copilots lies is in the abstraction level at which they operate: *tasks* (or more practically, localized code changes). In the real-world, software engineers are driven by *goals* (or more practically, GitHub issues), which sit at a much higher level of abstraction compared to tasks. However, solving GitHub issues is remarkably harder than accomplishing a single isolated tasks.

In this paper, we argue that copilots should be replaced with AI pair programmers. In our vision, an AI pair programmer behaves similarly to a senior pair programmer. The human engages with the AI pair programmer in a series of back-and-forth conversations to iteratively develop a solution to a problem (e.g., a GitHub issue). We also argue that an AI pair-programmer must be **goal-driven**, a **human partner** (not replacer), **SE-aware**, and a **self-learner**. Ultimately, our vision of an AI pair-programmer leads to a shift from AI-augmented SE to AI-transformed SE.

After presenting our vision of an AI pair programmer, we discuss key challenges that must be overcome before such a vision can be fully realized. These challenges were identified based on (i) surveys of academic and gray literature, (ii) in-depth discussions with industrial and academic leaders (e.g., during SEMLA 2023 [33], FM+SE Vision 2030 [25], and FM+SE Summit [26] events), (iii) meetings with our customers and our own internal development teams to understand their frustrations with copilots, and (iv) our practical experience with the research and development of novel technologies to enable AI pair programmers, including: a flexible multi-agent collaboration framework, agent-based cognitive architectures, and an efficient agent runtime that operates on heterogeneous hardware [27].

## 2 OUR VISION

### 2.1 Desired attributes of an AI pair programmer

The technological breakthroughs brought by FMs and FM-powered agents have the potential to take AI-assisted code generation to a whole new level of automation in which developers and AI become pair programmers.

To determine what the main attributes of these AI pair programmers should be, we reflect on two fundamental principles: (i) the ultimate goal of SE is to produce *great* software and (ii) SE is an inherently *socio-technical endeavor*.

• **Writing great software.** Great software must satisfy its customer (i.e., it must do what the customer wants it to do) [36]. Ultimately, code is just a means to an end. Therefore, AI pair programmers must be **goal-driven** (as opposed to task-driven). Being goal-driven also gives AI partners the freedom to leverage the ground-breaking planning capabilities of FMs without being limited by the confines of human cognition.

Second, great software is well-designed, well-coded, and easy to maintain, reuse, and extend [36]. Great software also fulfills nonfunctional requirements (a.k.a., quality attributes), such as performance and scalability. Therefore, given the depth and complexity of SE knowledge, AI pair programmers must be **SE-aware**. For instance, they must be able to reason about architecture trade-offs and design choices, as well as perform complex code changes that are not necessarily additive (e.g., refactoring).

• **SE as a socio-technical endeavor.** Software engineering is much about social interactions as it is about code. Following the pair programming paradigm, humans and AI collaborate as **partners**. The *driver* writes the code and the *observer* reviews each line of code as they are typed. The two programmers chat with each other and switch roles as convenient. As a human partner, the AI should exhibit adequate social traits, such as conversational intelligence (e.g., proactivity and communicability), social intelligence (e.g., manners and personalization), and personification (e.g., identity) [11]. As a practical example, AI partners should never overwhelm humans with information [49].

Finally, since we expect AI pair programmers to mimic humans, they should also learn from their mistakes based on feedback. Over time, AI pair programmers should also autonomously learn recurrent contextual information (e.g., project constraints regarding license usage) and human preferences (e.g., regarding coding style) to streamline communication. That is, AI pair programmers should also learn by themselves from self-reflection and cognitive exercises. We thus conclude that AI pair programmers should be **self-learners**.

## 2.2 Our vision of an AI pair programmer

Human and AI act as pair programmers to accomplish a goal (typically, a GitHub issue). As a first step, human and AI aligning on this goal. The human will communicate an initial version of such a goal and the AI will help the human to refine it. Next, pair programming effectively starts, and the solution is developed in an iterative fashion (humans and AI switch pair programming roles as convenient). The AI keeps track of the goal, refines it together with the human if necessary, and constantly *evaluates* progress toward that goal. We refer to this process as **Evaluation-Driven Delivery** (EDD). EDD draws inspiration from Test-Driven Development (TDD), which (i) compels developers to have a deeper and earlier understanding of the requirements and (ii) emphasizes the role of code as a means, not an end [6, 7]. In EDD, humans play a much bigger role and exert much more control comparatively to *AI-first* approaches like

ChatDev[44], MetaGPT [28], and Devin AI. In those approaches, AI takes the center stage and the human becomes the copilot.

We envision our AI pair programmer to be implemented with a set of FM-powered agents that collaborate among themselves (i.e., a multi-agent system). Different agents play different roles. We anticipate the need of the following agents:

• *Goal Alignment Agent.* This agent is responsible for helping the human figure out exactly what the goal should be and the context in which that goal should be delivered (e.g., project-specific constraints). In other words, this agent helps elicit requirements. It does so by (i) chatting with the human, (ii) refining their intents, and (iii) diligently asking clarifying questions (see Challenge 1). To increase productivity, humans converse with this agent as if they were talking to a human. That is, they do *not* write prompts. Instead, they express intents and clarify them (see Challenge 2).

• *Architecture & Design Agent.* This agent is capable of reasoning about architecture/design choices and their trade-offs. This agent has a deep understanding of SE, making good use of design patterns and best practices. In the early stages of a software product, this agent can collaborate with the developer to generate insightful input to team-wide architecture evaluation discussions (e.g., when the team is employing the *Architecture Tradeoff Analysis Method* (ATAM) [31])

• *Code Agent.* This agent understands, writes, and refactors code. It is the *doer* in EDD. This agent has a deep understanding of programming logic and testing, as well as technologies (target programming language, frameworks, and libraries). It is also capable of reasoning about its own code and producing associated repairs, ultimately reducing the need for human intervention. This agent is expected to be powered by FMs that understand the many facets of code (see Challenge 3), including static (e.g., structural dependencies) and dynamic (e.g., the values of variables in runtime and execution stacks) information.

• *Goal-Delivery Agent.* This agent tracks goal achievement. With the help of the *goal alignment* and *code agents*, it makes sure that (i) requirements are translated into tests, (ii) tests are adequately adapted as requirements change, and (iii) tests must eventually pass. By tracking progress toward current and past goals, this agent can be leveraged to promote human mentoring (see Challenge 4)

We clarify that we do not expect this vision of an AI pair programmer to be final, but rather to serve as a starting point for a broader discussion within the software engineering community. We plan to refine our vision as we (i) gain more experience designing these agents and their communication protocol, (ii) receive internal feedback from our development teams and customers, and (iii) receive external feedback from the broader SE community.

## 2.3 Theoretical underpinning

• **[Education] Bloom's 1984 Sigma Learning Problem.** According to Bloom [8], an average student performs two standard deviations better with one-on-one mentoring as compared to the conventional classroom learning environment. Knowledge is constantly shared between pair programmers. Hence, our expectation is that an AI pair programmer can mentor humans in a one-on-one fashion, leading to a significant improvement of their software engineering skills [41, 50] (see Challenge 4). Ultimately, we see the potential for AI pair programmers to take junior developers to a senior status much faster and with zero additional cost.

- **[Psychology] Theory of mind.** Theory of mind is a psychological concept that refers to one's ability to understand and attribute mental states to others [4]. At its core, theory of mind is about recognizing that other people have their own thoughts and perspectives that may be different from our own. Such a recognition is crucial for social interactions, as it helps us predict and interpret the behavior of others. Theory of mind is developed in early childhood and plays a critical role in our ability to communicate and form relationships. Microsoft shows that powerful models such as GPT-4 are able to develop theory of mind [9]. We thus expect that such a capability can be used to speed up goal alignment between humans and AI (see Challenge 1) as well as create tailored mentoring (see Challenge 4).
- **[Software engineering] The power of human-AI conversations.** Austin et al. [1] show that the percentage of software engineering tasks that can be solved with the help of an FM increases substantially when humans and AI engage in a back-and-forth conversation. The performance improvements achieved through these interaction rounds enable the use of smaller FMs that would otherwise be unable to compete with larger FMs. Our envisioned AI pair programming paradigm is strongly built on the idea that humans and AI will carry out back-and-forth conversations similarly to how a pair of humans would do it.

## 2.4 Our vision in context

As we showed in Section 1, current copilot solutions exhibit several limitations. JetBrains, the company behind popular IDEs such as PyCharm and IntelliJ, has recently partnered with OpenAI to provide a new AI assistant that provides functionalities that go significantly beyond code completion, such as code refactoring and suggestion of fixes to address issues. [40]. Such an endeavor makes it clear that the current code completion paradigm is too restrictive and that the industry has acknowledged that FMs can bring many more opportunities for boosting developer productivity.

We clarify that our vision of an AI pair programmer differs from that of Devin AI [13]. Devin AI is a state-of-the-art AI software engineer that can autonomously accomplish several software engineering tasks, including solving simpler GitHub issues. Currently, Devin AI can solve 13.86% of the GitHub issues in the SWE-bench benchmark [29] (approximately 3x more than GPT-4). This is a technically impressive result. However, we believe that such a technology still has to considerably evolve before it can be used in an enterprise context. For instance, the speed with which Devin AI can autonomously generate code makes it hard for developers to review that code and even maintain it in the future. If a critical bug is found in production (which is not unlikely [47]) and developers are little acquainted with the code (because it was mostly AI-generated), then maintenance becomes a clear bottleneck. Furthermore, architecture and design choices are all about trade-offs. Delegating these choices to an AI can lead to solutions that are simply inadequate in practice. Finally, Devin AI does not focus on goal alignment, which means that the resulting solution may not fulfill the original, true requirements. Therefore, while we are definitely positive and curious about the future of Devin AI (as well as its open source variations, such as OpenDevin [39], Devika [45], and SWE-agent [51]), we believe that it also comes with its own set of big challenges that must be overcome before it can be used in an enterprise context. Instead, we believe that AI pair programmers are much closer to adoption and that its challenges can be addressed in the short-term.

## 3 CHALLENGES

In this section, we introduce key challenges to realize our vision of AI software engineers. Overcoming these challenges requires novel technologies and a deep rethinking of how humans and AI interact.

### Challenge 1: Speeding up human-AI alignment

**Description.** An FM can only do what it is asked to do. That is, the output of an FM depends entirely on what one inquiries it with. Yet, humans have a limited ability to fully and clearly express their goals in one-shot using written text. As a consequence, human-AI alignment of goals becomes a challenge (e.g., humans may forget to communicate important aspects of a requirement). We note that this is a problem that will continue to exist no matter how powerful FMs may become, since it stems from an intrinsic human limitation. Humans are the bottleneck (not AI). To add to the challenge, natural language is inherently ambiguous [30].

**Our vision.** First, an AI pair programmer must be have excellent conversation skills (e.g., by using more powerful FMs) that can help humans refine their goals. Next, the AI must also ask humans for clarification *when needed*. Yet, finding the balance between asking too many clarifying questions and not asking enough is extremely hard. To mitigate this problem, AI pair programmers should have the ability to develop a theory of the mind of the human with whom they are interacting (see Section 2.3).

We draw the following analogy: the manner in which a task is described to a new development team member is typically more comprehensive compared to when that task is assigned to a senior team member. The senior team member already has assumptions in their head regarding the expectations of the task assigner (especially when they have had a history of interactions). Similarly, an AI pair programmer should learn from prior interactions with humans, such that goal details can be omitted, recurrent themes do not require repeated clarifications, and missing information can be inferred.

Designing effective mechanisms for enabling an AI pair programmer to develop theory of mind remains an open challenge. At the very least, such a mechanism would need to store prior conversations with the human, continuously summarize them, and inject them as appropriate into the context window of the underlying FM.

We emphasize, however, that theory of mind is not enough. Even with a perfect theory of someone's mind, there will likely be important contextual factors that the human will forget to mention or even consider (intrinsic human limitation). Hence, the AI pair programmer should constantly strive to refine the human's goal.

Finally, humans also need signals from the agents (e.g., confidence scores) to be able to judge whether those agents understood the communicated goal. Such signals are needed due to natural language ambiguity as well as inherent limitations of a text-based interaction design [43] (e.g., lack of visual cues such as facial expressions). Current conversational solutions like GitHub Copilot Chat [22] do not indicate by default how confident they are in their answers.

### Challenge 2: From prompts to natural inquiries

**Description.** An AI pair programmer should be able to understand inquiries at the same level that a senior software engineer would. For instance, developers should not have to worry about phrasing their inquiries as prompts that must be crafted using an unnatural lingo that implements some specific *prompt engineering* [15] technique (e.g., chain-of-thought [48]). After all, research shows that prompts are fragile [20] (i.e., even slight variations in a prompt can lead to very different outputs) and their effectiveness is model-dependent [12]. **Our vision.** In our vision, the burden of crafting an effective prompt should be on the AI (instead of humans). Although techniques such as Automatic Prompt Engineer [53], PromptBreeder [19] and DSPy [32] exist, they still require manual intervention (setup and/or programming). There is a need for a *prompt transpiler* technology that seamlessly takes a human inquiry and transforms it into an optimal prompt for a given model.

A promising research direction consists of gathering human feedback (e.g., thumbs up/down) for model responses then automatically using that feedback to improve the model's upcoming responses. The key idea is to store <instruction,response> pairs where response is of good quality (e.g., it received a thumbs up) and later on use those pairs to create few-shot examples in the system prompt. Over time, with a big enough database (e.g., built using crowdsourcing), the model learns how to appropriately answer questions. Such a functionality is being currently developed by LangChain [34] and highlights the importance of collecting semantic telemetry data to support model *evolution* and *self-learning*.

## Challenge 3: Cheaper and smarter code models

**Description.** An AI Pair Programmer must be able to fluently understand and write code (Section 2.2). While popular generalist FMs such as GPT-4.0 power popular copilot solutions, those models have key drawbacks when it comes to source code. First, they are oblivious of the rich nature of code by treating it as text and simply learning patterns during pretraining. Richer semantic information (e.g., code execution knowledge) is only partially learned. Second, they are typically too large, which means that they are expensive to train and use. Finally, the training data is not adequately curated, frequently violating copyright and license [23].

**Our Vision.** AI Pair Programmers should leverage Large Language Models for code (a.k.a. code LLMs). These models are specifically trained with source code, aiming to better capture code semantics compared to generalist LLMs. Code LLMs can be seen as *contextualized* models, in which special focus is given to the training data. The recent work of Lozhkov et al. [35] on StarCoder v2 show that curating the training data (e.g., selecting high quality sources, adhering to licenses) and applying careful preprocessing (e.g., ordering source code files per project and using an LLVM representation) can generate significantly smaller models with a performance that rivals that of much bigger models. For instance, the authors show that StarCoder2-3B outperforms StarCoderBase-15B, and that StarCoder2-15B outperforms CodeLlama-34B. Therefore, we believe that there are great research opportunities revolving around the creation of smaller (cheaper) yet effective code LLMs. In particular, we observe opportunities for creating multi-modal FMs that take into account both the static and the dynamic perspectives of code. For instance, Ding et al. [16] pretrain a multi-modal FM on a combination of source code and execution traces with the goal of teaching that FM complicated execution logic. The authors show promising results for clone retrieval and vulnerability detection.

## Challenge 4: Leveraging mentoring potential

**Description.** Humans are largely influenced by the people surrounding them. A famous quote often attributed to self-made millionaire and entrepreneur Jim Rohn says "You Are the Average of the Five People You Spend the Most Time With" [10]. Due to the socio-technical nature of SE [5, 46] (Section 2.1), developers are largely influenced by their mentors, team members, and peers. For instance, there is a vast literature discussing the role of mentors in the retention of newcomers in open-source software (OSS) [2, 14]. However, current copilots are not only vastly impersonal but also do not offer any kind of explicit mentoring.

**Our vision.** Social interactions play a big role in software engineering. In our vision, an AI pair programmer should not only code like a senior human engineer (or even better!), but also *mentor* like one. While research effort has focused on the former, we believe that the latter is equally important. After all, great software engineers are much more than good coders. More specifically, we believe that AI pair programmers should behave like emphatic team players that care about the technical growth and development of their peers.

The pair programming paradigm implicitly promotes mentoring by means of the back-and-forth conversations between the human and the AI (e.g., when the AI explains the trade-offs of a given design choice or how some specific framework works). However, we believe that there is ample room for explicit mentoring too. Similarly to how research in MSR (mining software repositories) has successfully created AI models that learn from history (e.g., defect prediction models [24, 52, 55]), we foresee great research opportunities in the design of AI pair programmers that (i) learn and reason from their prior interactions with humans, (ii) offer personalized technical mentoring to those humans (e.g., a personalized list of training resources at the end of a coding session), and (iii) tracks human SE skills development over time. Telemetry data is essential to enable these capabilities. In a nutshell, we believe that researchers should not only investigate how to make better AI pair programmers, but also investigate how to leverage those AI pair programmers to make better human software engineers.

## 4 CONCLUSION

Copilots are code completion systems at heart. While code completion is definitely useful in the acceleration of mundane tasks and exploration of new ideas [3], maintaining a large legacy system cannot be done on the basis of adding more and more code. Bootstrapping developers' productivity while simultaneously creating high-quality software requires a more shift from AI-augmented SE to AI-transformed SE.

In this paper, we propose the concept of an AI pair programmer that is **goal-driven**, a **human partner** (not replacer), **SE-aware**, and a **self-learner**. We also discuss key challenges that must be overcome to realize our vision of AI pair programmers. Fully tackling those challenges requires the design of novel technologies and techniques that touch on several knowledge areas (e.g., interaction design, software engineering, cognition, telemetry, and multi-agent collaboration). The list of challenges discussed in this paper is non-exhaustive and will be expanded in the future.

We hope that our vision and challenges will encourage deeper discussions in the software engineering community around the use of AI to bootstrap developer productivity.

## ACKNOWLEDGMENTS

The findings and opinions expressed in this paper are those of the authors and do not necessarily represent or reflect those of Huawei and/or its subsidiaries and affiliates. Moreover, our results do not in any way reflect the quality of Huawei's products.